\documentclass[journal]{IEEEtran}
\usepackage[T1]{fontenc}
\usepackage{cite}
\usepackage{amsmath,amssymb,amsfonts,bm}
\usepackage{algorithmic}
\usepackage{graphicx}
\usepackage{textcomp}
\usepackage{xcolor}
\usepackage{enumitem}
\usepackage{siunitx}
\usepackage{caption}
\usepackage{subcaption}
\usepackage{multirow}
\usepackage[ruled,vlined]{algorithm2e}

\DeclareMathOperator*{\argmin}{arg\,min}
\SetKwComment{Comment}{/* }{ */}
\SetKwInOut{STEPone}{\textbf{Step 1}}
\SetKwInOut{STEPtwo}{\textbf{Step 2}}
\SetKwInOut{STEPthree}{\textbf{Step 3}}
\SetKwInOut{STEPfour}{\textbf{Step 4}}
\SetKw{Return}{Return}

\SetCommentSty{mycommfont}
\usepackage{mathtools}
\makeatletter
\let\old@ps@IEEEtitlepagestyle\ps@IEEEtitlepagestyle
\def\confheader#1{%
    \def\ps@IEEEtitlepagestyle{%
        \old@ps@IEEEtitlepagestyle%
        \def\@oddhead{\strut\hfill#1\hfill\strut}%
        \def\@evenhead{\strut\hfill#1\hfill\strut}%
    }%
    \ps@headings%
}
\makeatother
\confheader{\parbox{20cm}{\textbf{\textit{Accepted to IEEE Belgrade PowerTech, 2023}}}}

\begin{document}

\renewcommand{\figurename}{Fig.}
\renewcommand{\tablename}{TABLE}

\title{Regularised Learning with Selected Physics for Power System Dynamics}

\author{Haiwei Xie~\IEEEmembership{Graduate Student Member,~IEEE}, Federica Bellizio~\IEEEmembership{Member,~IEEE}, Jochen L. Cremer~\IEEEmembership{Member,~IEEE}, Goran Strbac~\IEEEmembership{Member,~IEEE}  \vspace{-0.25em}
\thanks{H. Xie and J. L. Cremer are with Department of Electrical Sustainable Energy, TU Delft, The Netherlands (e-mail of corresponding author: j.l.cremer@tudelft.nl).
F. Bellizio is with the Urban Energy Systems Laboratory, Empa, Switzerland.
G. Strbac is with the Department of Electrical and Electronic Engineering, Imperial College London, UK.}
}

\maketitle

\begin{abstract}
Due to the increasing system stability issues caused by the technological revolutions of power system equipment, the assessment of the dynamic security of the systems for changing operating conditions (OCs) is nowadays crucial. To address the computational time problem of conventional dynamic security assessment tools, many machine learning (ML) approaches have been proposed and well-studied in this context. However, these learned models only rely on data, and thus miss resourceful information offered by the physical system. To this end, this paper focuses on combining the power system dynamical model together with the conventional ML. Going beyond the classic Physics Informed Neural Networks (PINNs), this paper proposes Selected Physics Informed Neural Networks (SPINNs) to predict the system dynamics for varying OCs. A two-level structure of feed-forward NNs is proposed, where the first NN predicts the generator bus rotor angles (system states) and the second NN learns to adapt to varying OCs. We show a case study on an IEEE-9 bus system that considering selected physics in model training reduces the amount of needed training data. Moreover, the trained model effectively predicted long-term dynamics that were beyond the time scale of the collected training dataset (extrapolation). 
\end{abstract}


\section{Introduction}\label{sec1}
To ensure power system secure operation, it is critical to analyse the system dynamic security including the assessment of various stability phenomena during faults (such as rotor angle, frequency, or voltage stability~\cite{kundur2004definition}~\cite{panciatici2012operating}). Under higher operational uncertainty brought by the increasing renewable energy sources, the development of real-time Dynamic Security Assessment (DSA) tools are high on the system operator’s agendas~\cite{Ent17}. One challenge towards real-time DSA is that the conventional DSA tools need high computation time as they rely on slow numerical integration of a dynamical model described with Ordinary Differential Equations (ODEs), where a fault is simulated as a discrete event~\cite{Har07,Pop17}. 

Extensive research efforts have aimed at making DSA real-time applicable, mainly falling in two categories. The first category is to simplify the calculated models. For example, \cite{Pav12} introduced a single-machine equivalent of the dynamical models only considering the critical generator, and \cite{Vu15} evaluated theory on energy functions (Lyapunov functions). However, these methods are limited when generalized to large-interconnected power systems due to the strong assumptions they place on the system model itself. The second category is to use ML approaches to predict dynamic security. The approaches in this category have received great attention over the last three decades~\cite{duchesne2020recent, Shi20}. The main underlying idea is, to train a model in the offline stage based on obtained datasets including features of operating conditions and their stability label, and apply the trained model online to predict the security status of the real-time operation. Many models have been studied in this context, such as Decision Trees for interpretable predictions~\cite{Weh93,Bug21,cremer_optimization-based_2019,hou2020sparse} or Artificial Neural Networks (ANNs) with deep learning~\cite{dong2013using,Zha21}. These research work have yielded fruitful results, demonstrating that ML holds great potential for real-time DSA by making use of the repetitive nature of the operating conditions in power system operation.

Nonetheless, it is also noticed by the community that these ML approaches have limitations caused by their fully data-based nature, e.g. high dependency on data quality and quantity, low interpretability, physical consistency/feasibility of output results. This encourages researchers to apply PINNs~\cite{huang2022applications} that aim at incorporating rich power system theory into ML models. The PINN is a supervised machine learning model~\cite{raissi_physics-informed_2019}, formulating the physical model into the training loss function such that the model can provide a more physical-consistent solution with less training data. It has been successfully applied to system identification~\cite{stiasny_physics-informed_2020}, predict dynamics of individual generators~\cite{misyris2020physics} and for the assessment of system-wide transient stability~\cite{stiasny2021transient}. However, a research gap is that these existing PINN-based approaches formulated only one unchanged system dynamical model into the training loss function. When the system OC changes, accordingly the system dynamical model changes with the nonlinear updates of ODE equation parameters, and then the PINN models are invalid. 

This paper bridges the identified gap by proposing Selected-Physics-Informed Neural Network (SPINN) that regularises the training of the PINN approach with a subset of equations scalable to new operating conditions, in the context of learning power system dynamics. The main contributions of this paper are:
\begin{enumerate}
    \item We propose the SPINN with a two-level structure, to regularize its learning with selected physics scalable to varying operating conditions.
    \item We put forward a sampling strategy for SPINN, which can largely reduce the model’s data dependency and improve the dynamic prediction for longer time period without any measurement data.
\end{enumerate}

Through experiments on a IEEE-9 bus system, we show the prediction performance, operating conditions scalability, extrapolation capabilities, and computation time of SPINN in comparison with vanilla NN and PINN. 

The rest of the paper is structured as follows. The power system dynamics under varying operating conditions are introduced in Section II. Thereafter, Section III introduces the proposed SPINNs for learning power system dynamics. Subsequently, the case study is in Section IV and conclusions in Section V. 

\section{Power system dynamics under varying operation conditions}\label{sec2}
\begin{figure}
    \centering
    \includegraphics[width=1\linewidth]{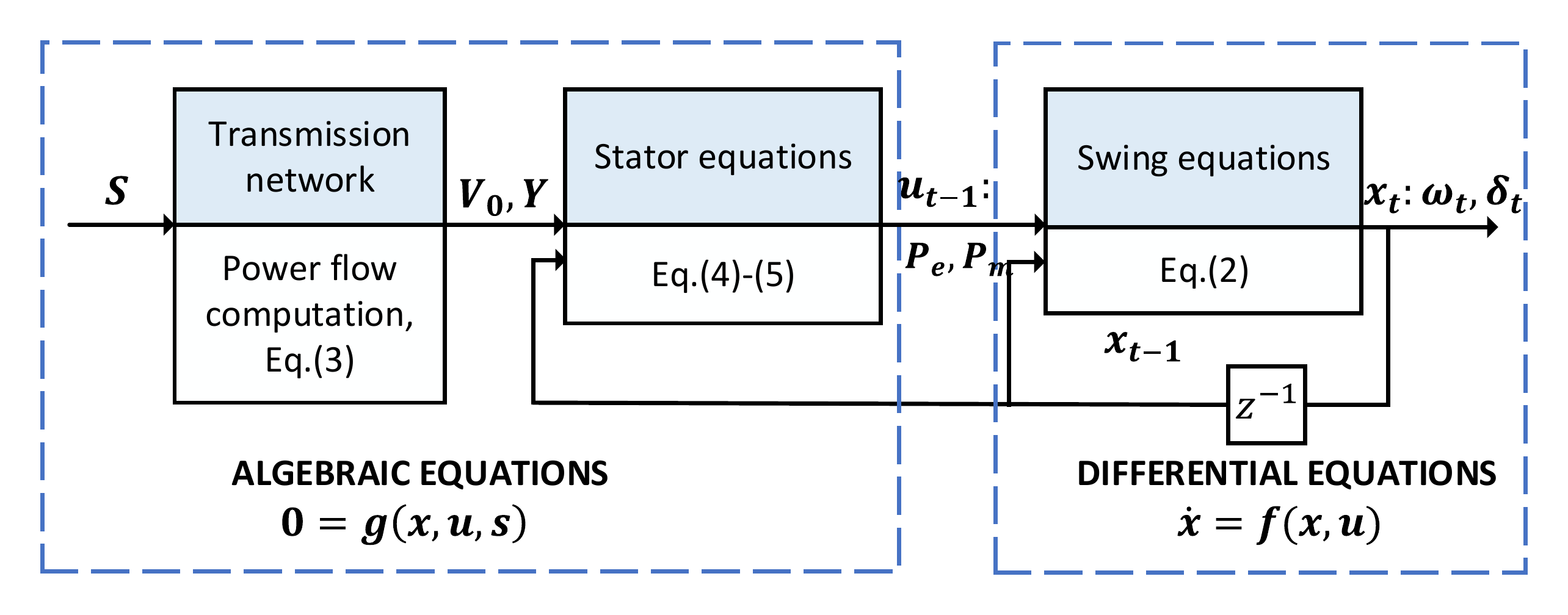}
    \caption{Power system dynamics computation.}
    \label{fig1}
\end{figure}

In this section, we briefly review the power system dynamical model and its computation (as in Fig.~\ref{fig1}), stressing especially its computation under varying operational conditions $\tilde{\bm{s}}$.

The power system dynamics~\cite{stott1979power} can be modelled as a set of differential equations parameterized by $\Theta$, 
\begin{subequations}
\begin{equation}
  \dot{\bm{x}}=\bm{f}_{\Theta}(\bm{x},\bm{u}), 
  \label{eq1a}
\end{equation}
and algebraic equations
\begin{equation}
 \bm{0}=\bm{g}(\bm{x},\bm{u},\tilde{\bm{s}}),
 \label{eq1b}
\end{equation}
\label{eq1}
\end{subequations}
\noindent where $\tilde{\bm{s}}$ is the vector featuring operating conditions, usually represented by the load vector. $\bm{x}$ is the states vector, $\bm{u}$ is the inputs vector, which we will specify later.

In a $m$-bus power system with $n$ machines ($n<m$), \eqref{eq1a} includes the equations of motion with regard to each generator's rotor angle $\delta_k$ and deviating frequency $\omega_k$,
\begin{subequations}
\begin{equation}
  \frac{\mathrm{d} \omega_{k} }{\mathrm{d} t}= \frac{1}{m_k}(-d_k\omega_k-P_{elec,k}+P_{mech,k}),
\end{equation}    
\begin{equation}
 \frac{\mathrm{d} \delta_{k} }{\mathrm{d} t}=\omega_k,
\end{equation}
\label{eq2}
\end{subequations}
parameterized by $m_{k}$ the inertia constant of the $k$th machine, and $d_{k}$ the mechanical damping of the $k$th machine proportional to frequency deviation. Therefore, the system states $\bm{x}$ are a vector including all $\delta_k$ and $\omega_k$, $k=1,2,...n$. The inputs $\bm{u}$ are a vector including all mechanical power inputs $P_{mech,k}$, and the electrical power $P_{elec,k}$.

The inputs $\bm{u}$ are computed with the algebraic equations~\eqref{eq1b}, a set of implicit functions w.r.t $\bm{u}$, $\bm{s}$ and $\bm{x}$. The computation process consists of i) calculate reduced-order admittance matrix $\bm{Y}_{r}$ (kron-reduction~\cite{dorfler2012kron}) and pre-fault voltage on each generator bus $\bm{V}_0$ (power flow computation), ii) calculate the mechanical power and electrical power respectively with stator equations. The detailed equations are as follows
 
 \begin{subequations}
     \begin{equation}
         \bm{Y}=\begin{bmatrix}
    \bm{Y}_e^{r*r} & \bm{Y}_c^{r*n}\\
    \bm{Y}_b^{n*r} & \bm{Y}_g^{n*n}
    \end{bmatrix},
     \end{equation}
\begin{equation}
     \bm{Y}_{r}=\bm{Y}_{g}-\bm{Y}_{b}\bm{Y}_{e}^{-1}\bm{Y}_{c},
\end{equation}
\label{eq3}
 \end{subequations}
 where $r=m-n$ and corresponds to the buses without generators, $\bm{Y}$ is the original full admittance matrix  with $m$ buses. 

\begin{subequations}
    \begin{equation}
        \bm{V}_{g}=\|\bm{V}_{0}\|e^{j\bm{\delta}},
        \label{eq4a}
    \end{equation}
    \begin{equation}
        \bm{I}_{g}=\bm{Y}_{r}\bm{V}_{g},
        \label{eq4b}
    \end{equation}
    \begin{equation}
         \bm{P_{elec}}=Real(\overline{\bm{I}_{g}}\odot \bm{V}_{g})
         \label{eq4c}
    \end{equation}
    \label{eq4}
\end{subequations}
where $\odot$ represents the Hadamard product of matrices, $\bm{V}_{g},\bm{I}_{g}$ are the voltage and current at generation buses, $\overline{\bm{I}_{g}}$ is the complex conjugate of $\bm{I}_{g}$.

$\bm{P_{mech}}$ is equal to the prefault electeical power $\bm{P_{elec0}}$ as the system is assumed to be stable before the fault occurs.
\begin{equation}
     \bm{P_{mech}}=\bm{P_{elec0}}=Real(\overline{\bm{Y}_{r}\bm{V}_{0}}\odot \bm{V}_{0})
     \label{eq5}
\end{equation}

Eq~\eqref{eq3}-\eqref{eq5} illustrate that the algebraic equations set $\bm{g}$ is implicit functions with regard to not only $\bm{x}$, $\bm{u}$, but also $\tilde{\bm{s}}$. On the other hand, when the generator parameters $m_k$, $d_k$ are set and known, the differential equations set $\bm{f}$ formulates only a relationship between $\bm{x}$ and $\bm{u}$, thus it is the same for all varying OCs.

\section{Learning power system dynamics with SPINNs}\label{sec3}
\begin{figure*}[t!]
    \centering
    \includegraphics[width=0.8\linewidth]{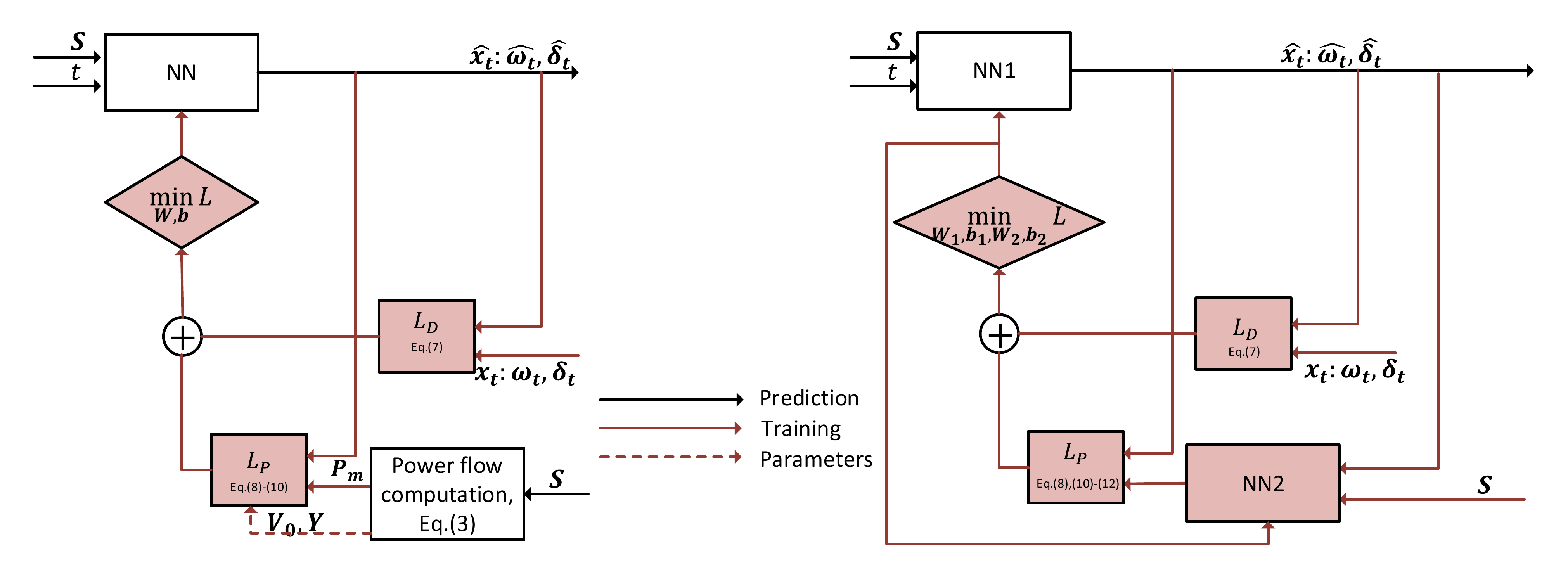}
    \caption{The scheme of PINNs (left) and SPINNs (right).}
    \label{fig:2}
    \vspace{-0.5cm}
\end{figure*}
In this section, we propose the SPINN to learn the power system dynamics described above. We first introduce the workflow of supervised ML based model for learning power system dynamics. Then we introduce our proposed structure of SPINN and the proposed training data sampling strategy. Finally, we compare the SPINN model with the PINN model and show the improvements.
\subsection{Supervised ML for power system dynamics prediction}\label{sec3.1}
Supervised learning can train the parameters of a ML model based on a given training dataset (offline stage), such that the model learns the relationship between inputs and outputs during the training. The trained model is able to be applied to directly predict the power system dynamics given inputs (online stage). 

The workflow of supervised ML for power system dynamics prediction is summarized in Algorithm.~\ref{alg:1}. Firstly, the training dataset should be prepared (discussed in~\ref{sec3.3}). The training data usually has a structure of inputs $\bm{I}$ constructed with $[\bm{s},\bm{t}]$ (i.e. $I^s_t=[\bm{s},\bm{t}]$ where $\bm{s}$ is the operating condition and $t$ is the time point), and outputs $\bm{X}$ constructed with the corresponding $\bm{x}^{s}_t$. The model is subsequently trained with the constructed data by minimizing the defined loss function. The loss function usually represents the deviation of the prediction $\hat{\bm{x}}^{s}_t$ from the true label $\bm{x}^{s}_t$. The formulation of loss function is choice-specified, e.g. mean square error (MSE)~\cite{stiasny_physics-informed_2020}.


After the training loss is minimized to fall into a given bound, the model is said to be trained and can be tested on testing data or used later in practice, by providing a prediction of states $\{\bm{x}\}$ given a $I^s_t$. The trajectory of a test OC $\bm{s}$ can be computed by giving the inputs $\{I^s_t\}_{t=1:T}$ and obtain outputs $\{\hat{\bm{x}}^s_t\}_{t=1:T}.$ 
\begin{algorithm}
\caption{Supervised ML for power system dynamics prediction}\label{alg:1}
\KwIn{\\(1) A defined model (e.g. NN) $f$ with its trainable parameters $\bm{W}, \bm{b}$;\\
(2) A given dataset $\bm{D}$ (inputs $\bm{I}$ and output $\bm{X}$) and be splitted into two, $\bm{D}^{tr}$ (inputs $\bm{I}^{tr}$ and output $\bm{X}^{tr}$) for training and $\bm{D}^{va}$ (inputs $\bm{I}^{va}$ and output $\bm{X}^{va}$) for validation.\\
(3) A testing operating conditions dataset $\bm{S}$ whose dynamics over time period $0-T^*$ need to be predicted.}
\KwOut{\\
(1) The trained model with parameters $\bm{W}^*, \bm{b}^*$;\\
(2) The predicted dynamics $\{\hat{\bm{x}}^s_t\}_{0\leq t\leq T^*}$ for all $\bm{s} \in \bm{S}$.}
\STEPone{Initialize the model parameters $[\bm{W}, \bm{b}]\gets [\bm{W}_0, \bm{b}_0]$.}
\STEPtwo{Define the training loss function $L(\hat{\bm{X}},\bm{X})$.}
\STEPthree{}
\While{$L(\hat{\bm{X}}^{va},\bm{X}^{va})> \epsilon$}{
$\hat{\bm{X}}^{tr}=f_{\bm{W}, \bm{b}}(\bm{I}^{tr})$;\\
$[\bm{W}, \bm{b}]\gets \argmin_{\bm{W}, \bm{b}}{L(\hat{\bm{X}}^{tr},\bm{X}^{tr})}$;\\
$\hat{\bm{X}}^{va}=f_{\bm{W}, \bm{b}}(\bm{I}^{va})$.
}
$[\bm{W}^*, \bm{b}^*]\gets [\bm{W}, \bm{b}].$\\
\STEPfour{}
\For{$0 \leq t \leq T^*$ and $\bm{s} \in \bm{S}$}{$\hat{\bm{x}_t^s}=f_{\bm{W^*}, \bm{b^*}}(I^{s}_t)$.}
\Return $\bm{W}^*, \bm{b}^*, \{\hat{\bm{x}_t^s}\}_{0\leq t\leq T^*}$
\end{algorithm}


\subsection{Regularised learning with selected physics}\label{sec3.2}

Following the above framework, this section explains in detail the design of the model, such that the model can be trained to provide accurate prediction of system states with less training data. 

We consider incorporating physics into ML framework, by regularizing the training of NN model with an additional term $L_P$ in its training loss function $L$, quantifying the deviation of the model prediction from the true system dynamics equations~\eqref{eq1},
\begin{equation}
    L=L_D+L_P,
    \label{eq6}
\end{equation}
where, if MSE is used,
\begin{equation}
    L_D=\frac{1}{N_l}\sum_{j=1:N_l}{(\hat{\bm{x}}_j-\bm{x}_j)^2} 
    \label{eq7}
\end{equation}
with $N_l$ the cardinality of the labelled dataset, and
\begin{equation}
    L_P=\frac{1}{N_c}\sum_{j=1:N_c}{(\hat{\bm{p}}_j-\bm{p}_j)^2},
    \label{eq12}
\end{equation}
\begin{equation}
    \hat{\bm{p}}_j=\dot{\hat{\bm{x}}}_j-\bm{f}(\hat{\bm{x}}_j,\bm{u}_{s}), \forall s,
    \label{eq8}
\end{equation}
\begin{equation}
    \bm{p}_j=\dot{\bm{x}}_j-\bm{f}(\bm{x}_j,\bm{u}_{s})\equiv \bm{0}, \forall s.
    \label{eq9}
\end{equation}
In the equations, $N_c$ is the number of samples used for computing physical deviation, also called as collocation points. $\hat{\bm{p}}_j$ is the computed physical deviation of the prediction $\bm{x}_j$, and $\bm{p}_j$ is the expected physical deviation of the true label, which is naturally $\bm{0}$. Thus the collocation points do not need to be labelled as the previously labelled data for computing $L_D$. $\bm{u}_s$ is the corresponding inputs under the operating condition $s$ of $\bm{x}_j$.

However, as we covered above, $\bm{u}_s$ is computed with~\eqref{eq1b}, and needs to be recomputed for each $s$ and $\bm{x}$. This is not an ideal case, as the training dataset contains points under a large number of different operating conditions. Updating the inputs $\bm{u}_s$ for each $s$ and $\bm{x}$ will bring about unnecessary issues for formulating the training loss function, and add additional computation burden. 

In response to these issues, we propose a method to only inform the NN training with a selected set of equations/ physics which are scalable to different OCs, to avoid the need of updating inputs with algebraic functions.

The key observation is, if $u_s$ is obtained somehow, the equation~\eqref{eq1a} itself is scalable to different OCs. Therefore, in SPINN, we propose to use another NN as a complementary, to approximate $u_s$ from $s$ and $\hat{\bm{x}}$ (the workflow of SPINN as shown in Fig.~\ref{fig:2}) 
\begin{equation}
    \hat{\bm{u}}_s=f_{NN2}(\hat{\bm{x}}_j,s),
    \label{eq10}
\end{equation}
where $f_{NN2}$ is the mathematical function representing NN2. With the approximated $\hat{\bm{u}}_s$, we can naturally substitute~\eqref{eq8} with 
\begin{equation}
    \hat{\bm{p}}_j=\dot{\hat{\bm{x}}}_j-\bm{f}(\hat{\bm{x}}_j,\hat{\bm{u}}_{s}), \forall s.
    \label{eq11}
\end{equation}
As shown in Fig.~\ref{fig:2}, the NN1 and NN2 in SPINN are trained together with~\eqref{eq6}, regularizing the outputs of NN1 $\hat{\bm{x}}$ not only to be close to the labels in the training dataset, but also to be physics continuity, i.e., following the system dynamic equations. Another benefit of the proposed SPINN is enabling us to include $N_c$ collocation points for physical deviation verification in the training other than only the original training dataset with $N_l$ samples, as we will explain in the next subsection. 

\subsection{A sampling strategy for training data}~\label{sec3.3}
\begin{figure}
    \centering
    \includegraphics[width=1.02\linewidth]{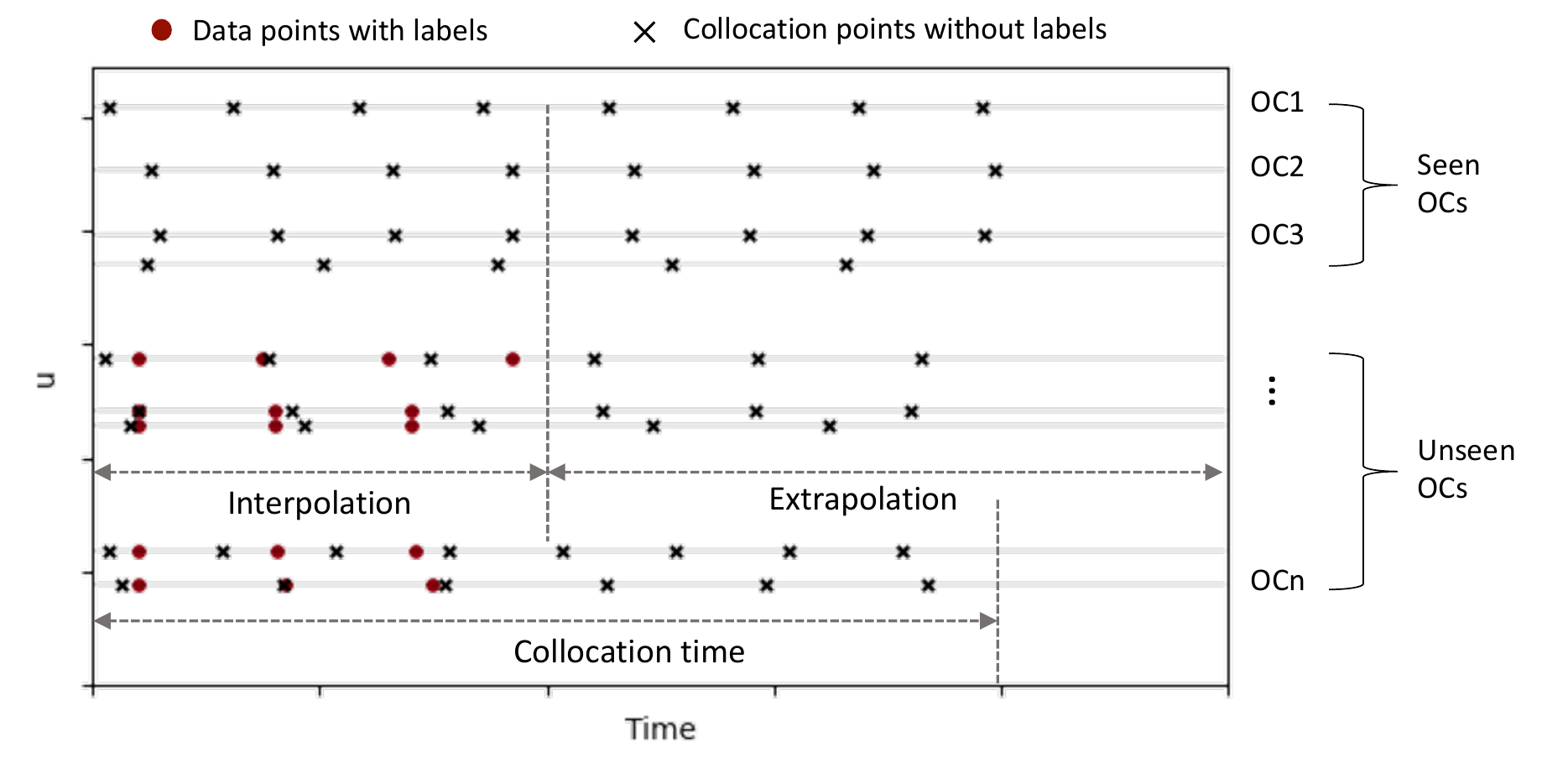}
    \caption{A scheme of the training data sampling strategy of SPINNs.}
    \label{fig3}
\end{figure}

We have shown above that there are two categories of training data considered in the SPINN training: $N_l$ labelled data samples for computing prediction error $L_D$, and $N_c$ collocation points for computing physical loss $L_P$, as visualized in Fig.~\ref{fig3}. The preparation of the training datasets for SPINNs follows the procedures in Algorithm~\ref{alg:two}.

For labelled data samples, their labels $\bm{x}$ come from simulation results. The procedure is to firstly obtain a set of seen operating conditions $\bm{S}=\{s\}$. Secondly, for each operating condition $s$, $n_s$ time points $\{t_s\}$ are sampled, and subsequently the dynamics at those time points $\{\bm{x}^s_t\}$ are simulated with an ODE solver, using system equations~\eqref{eq1}. One data sample has one inputs vector $I^s_t=[\bm{s},\bm{t}]$ and one outputs vector $\bm{x}^s_t$. The operating conditions in set $\{s\}$ are considered as seen OCs, as their dynamics are simulated before training.

In contrast, for collocation points, as clarified in~\eqref{eq9}, the computation of physical loss $L_P$ does not need labels $\bm{x}_j$ or $\dot{\bm{x}}_j$, as $\bm{p}_j \equiv 0$. Therefore, for these collocation points, only $\{[\bm{s},\bm{t}_p]\}$ need to be sampled in the input domain. After feeding them into the SPINN, the output $\hat{\bm{x}}$ and $\dot{\hat{\bm{x}}}$ will be used to compute the physical deviation, using~\eqref{eq8}-\eqref{eq11}. 

Withstanding the seen OCs are included to sample collocation points, we can extend the samples $\bm{S}$ to more unseen OCs, to improve the model's scalability to more OCs. Moreover, in order to improve the prediction accuracy of SPINN for longer time, the sampling of $\{t_p\}$ can be extended to extrapolation time which are beyond the time period of simulated data.

\begin{algorithm}
\caption{A sampling strategy for training data}\label{alg:two}
\KwIn{\\(1) The required cardinality for labelled dataset $N_l$ and for collocation (unlabelled) dataset $N_c$;\\
(2) Electrical network information and the given fault information;\\
(3) A set of seen OCs $\{\bm{s}\}_{seen}$ with cardinality $N_S$}
\KwOut{\\
(1) The labelled dataset $D_l$;\\
(2) The collocation dataset $D_c$.}
\STEPone{Allocate $N_l$ sampling points to each seen OCs (e.g. on average), i.e. $\sum_{s\in \{\bm{s}\}_{seen}}{n_s}=N_l$.}
\tcc{$n_s$ are the samples under OC $s$}
\STEPtwo{}
$D_l=\emptyset$;\\
\For{$s \in \bm{S}$}{sample $n_s$ time points $\{t_s\}$ in $[0,T_{in}]$;\\
get $\bm{x}^s_t$ by solving~\eqref{eq1} with an ODE solver;\\
$D_l\gets D_l \cup \{s,t,\bm{x}^s_t|\forall t \in \{t_s\}\}$.
}
\STEPthree{}
Sample a set of unseen OCs $\{\bm{s}\}_{unseen}$ and the whole OC set is $\{\bm{s}\}=\{\bm{s}\}_{seen}\cup \{\bm{s}\}_{unseen}$;\\
$D_c=\emptyset$;\\
Allocate $N_c$ sampling points to all OCs (e.g. on average), i.e. $\sum_{s\in \{\bm{s}\}}{n_s}=N_c$;\\
\For{$s \in \bm{S}$}{sample $n_s$ time points $\{t_p\}$ in $[0,T_{ex}]$;\\
$D_c\gets D_c \cup \{s,t,0|\forall t \in \{t_p\}\}$.
}
\Return $D_l, D_c$
\end{algorithm}

\subsection{Comparison with PINNs}\label{sec3.4}
\begin{figure*}[!t]
    \centering
    \includegraphics[width=1\linewidth]{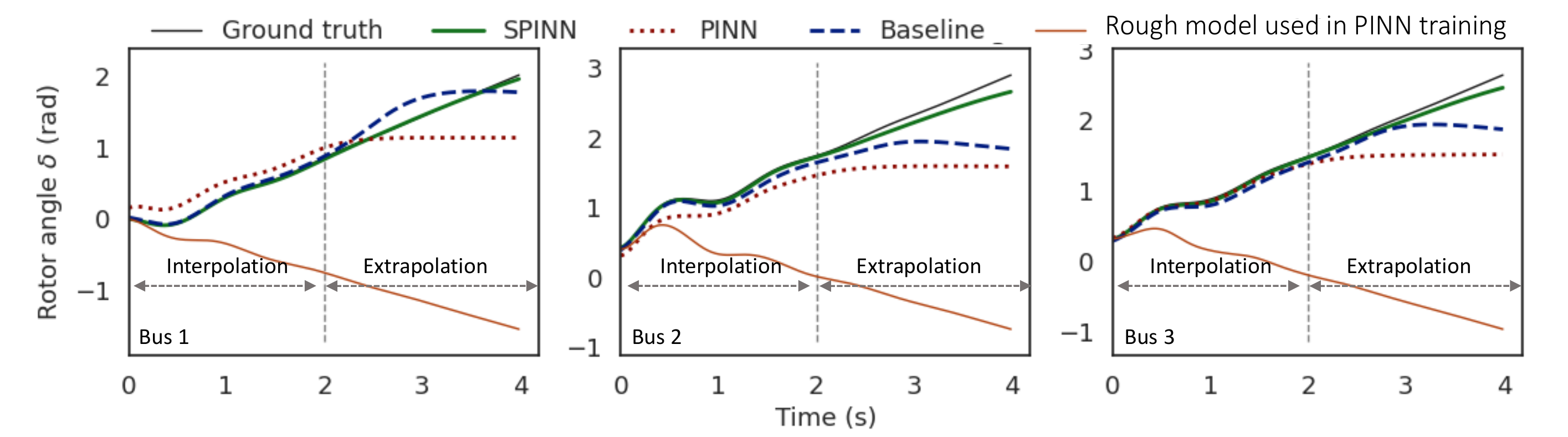}
    \caption{The prediction results for a new OC (not in training datasets) using the SPINN, PINN and baseline for three generators.}
    \label{fig:fig4.2.1}
    \vspace{-0.5cm}
\end{figure*}
The main differences between SPINNs and PINNs are summarized as follows and are visualised in Fig.~\ref{fig:2}.

i) Different from SPINNs that assume the system inputs $\bm{u}$ consist of both the mechanical power and electrical power of each generator $\bm{P_{mech}}$ and $\bm{P_{elec}}$, PINNs assume that $\bm{u}$ consist of only $\bm{P_{mech}}$, thus~\eqref{eq4} are also included in the computation of $L_P$ in order to describe $\bm{P_{elec}}$ as functions of $\bm{x}$. In contrast, in SPINNs, only~\eqref{eq2} (or~\eqref{eq1a}) are formalized in the loss function and an extra NN2 is used to approximate~\eqref{eq3}-\eqref{eq4} (or~\eqref{eq1b}) to estimate $\bm{u}$ for each OC $s$ and states $\bm{x}$.

ii) As shown in section~\ref{sec2},~\eqref{eq4} are a set of algebraic equations with regard to $\bm{x}$ but parameterized with $\bm{s}$. In order to formalize them, it is required to additionally compute all the parameters of~\eqref{eq4} such as $\bm{V_0}$ and $\bm{Y_r}$ from given $\bm{s}$. This stage increases the difficulty of formalizing ~\eqref{eq4} as these parameters are solved from implicit functions such as power flow computation. Once the OC $\bm{s}$ is given,~\eqref{eq4a} and~\eqref{eq4b} are solved to obtain the parameters which are needed in the formalization of~\eqref{eq4c}. Therefore, PINNs are trained only using the equations developed from one OC, which means it is not scalable to multiple OCs in nature. For each different OC, the whole process in algorithm~\ref{alg:1} needs to be go through again. In contrast, SPINNs only include~\eqref{eq2} in the computation of training loss $L_p$ which are a set of differential equations scalable to all OCs. Therefore, SPINNs are trained to be able to predict dynamics under multiple OCs.

\section{Case Study}\label{sec4}
In this section, we firstly study the issues of PINNs when learning power system dynamics under varying operating conditions. Subsequently, we investigate SPINNs in terms of the scalability to varying operating conditions and their extrapolation capabilities. Finally, we study the computational performance of SPINNs.
\subsection{Test system and assumptions}\label{sec4.1}
The case study used the IEEE 9-bus test system as parameterised in~\cite{garces_2016} having $3$ generators and $3$ loads. 
The training data involved $N_{s}=100$ load scenarios, each defineing a different initial OC. These initial OCs were generated by sampling the active and reactive loads on three buses from three independent Gaussian distributions centered at $1.25+j0.45$, $0.95+j0.25$ and $1.0+j0.3$, all in p.u., and had standard deviations around $0.5$, respectively.

The time-domain analysis of security involved a short-circuit fault at bus $7$, which was cleared after $0.1$s. 
The time-domain simulations were carried out in Matlab R$2020$a using the sixth stage-fifth order Runge-Kutta method (ode$45$ function). All studies were carried on a standard laptop machine with 4 cores and 16GB RAM. The post-fault transients were collected for a time period of $5$s and the transient data was recorded for equidistant times (every $0.01$s). 
The interpolation time was $[0, 2]$s and the extrapolation was $[2, 4]$s. During the NN training only the data from interpolation time was used.

The two NN structures in the SPINN model had $2$ hidden layers each with $128$ neurons. 
The activation functions $\sigma$ were selected as tanh and the Xavier's normal method was used to initialize the NN parameters $\bm{W}_0, \bm{b}_0$. The models $\mathit{NN1}$ and $\mathit{NN2}$ were all trained for $2000$ epochs with a varying batch size to improve the training efficiency.
The relative MSE of rotor angle predictions was used to evaluate the model accuracy on the testing dataset.

\subsection{Issue of PINNs for power system dynamics}\label{sec4.2}
This case studied the issue of PINNs of not scaling well to unseen OCs as the dynamic model incorporated in the model training is OC-specific. The dynamic model from an arbitrary selected OC1 (the initial load scenario) was used. A PINN was trained using the dynamics of OC1 and the data of $70$ other OCs. Subsequently, the PINN was tested on $30$ new OCs resulting in a very high prediction error of $31\%$. Fig.~\ref{fig:fig4.2.1} shows the prediction of the PINN for one arbitrarily selected testing OC (out of the $30$ OCs). This figure also shows the corresponding ground truth from time-domain simulations, the time-domain simulation when using the assumed incorrect dynamical model, and the improved SPINN performance.
 
The issue of PINNs can be seen when comparing with the ground truth showing clear differences with very large inaccuracies in all three generators.
This analysis demonstrates PINNs had very high prediction error (in average of $31\%$) for uncertain OCs, thus it is important to consider the correct dynamic model relevant for the OC.

\subsection{SPINNs under varying operating conditions}\label{sec4.3}
\begin{figure}[t]
    \centering
    \includegraphics[width=0.8\linewidth]{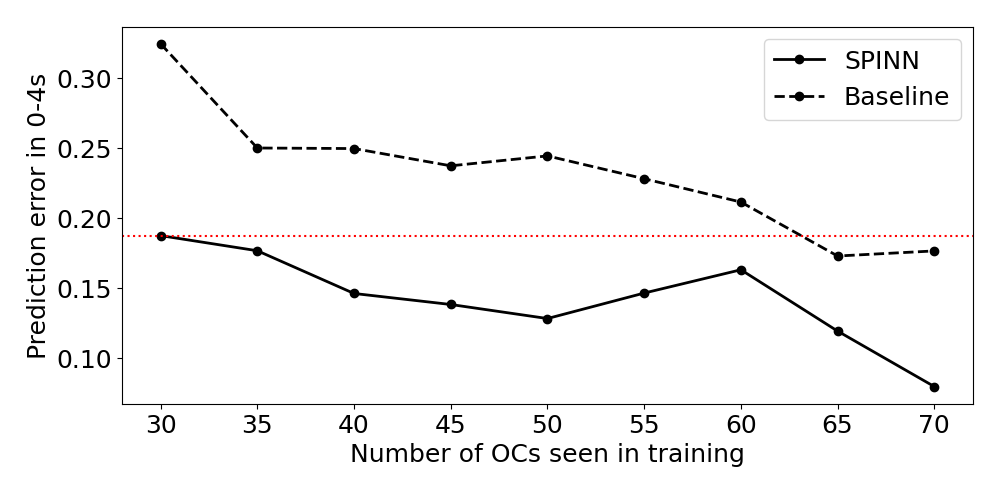}
    \caption{The prediction error using SPINNs and baseline models when varying the number of training OCs.}
    \label{fig:fig4.3.1}
\end{figure}
This case studied the scalability of SPINNs under varying operating conditions addressing the issue of PINNs (studied just before). Two studies were carried out, one to demonstrate that SPINNs effectively regularise with physics, the other to study the impact of the amount of training data on the performance to uncertain OCs. 

The first study used the same settings as before. The results are in Fig.~\ref{fig:fig4.2.1}. SPINNs outperformed all baseline models. The prediction accuracy for all the 30 OCs improved by around $10\%$ when trained with the same settings as PINNs. The same setting considered the number of data, seen OCs and unseen OCs (collocation points). These results show that the physical regularisation supported conforming better to the physics and improving prediction performance in SPINNs. 

The second study investigated SPINNs for varying the number of `seen' training data  $N_{OC}=\{50,55,60,65,70\}$. The SPINNs sampling strategy was that $D1=60$ data points were sampled in the interpolation time $[0, 2]$s for each training OC, and $D2=200$ collocation points were sampled from all $100$ OCs in the extrapolation time interval $[0, 4]$s. 
A baseline model considered no collocation samples $D2=0$. The results of the study are in Fig.~\ref{fig:fig4.3.1}. The average prediction error across the total time $[0-4]$s for $30$ unseen, new OCs reduced with increasing $N_{OC}$ in the two models, the SPINN model and the baseline model. The training data supported the models to better learn the mapping from inputs to system states. The key advantage is that SPINN required less data for the same error. For instance, when SPINNS were trained with only $30$ OCs this results in equivalent error when training the baseline model with $60$ OCs, reducing the needed data by $50\%$. An additional finding is that SPINNs show a higher prediction stability with less data dependency which is a result from the selected physics. E.g., when the number of training OCs was reduced, the error of SPINNs increased less (from $8\%$ to $19\%$) than in normal NNs (from $18\%$ to $32\%$). 

\subsection{SPINNs training data sampling strategies}\label{sec4.4}
\begin{figure}
    \centering
    \includegraphics[width=1\linewidth]{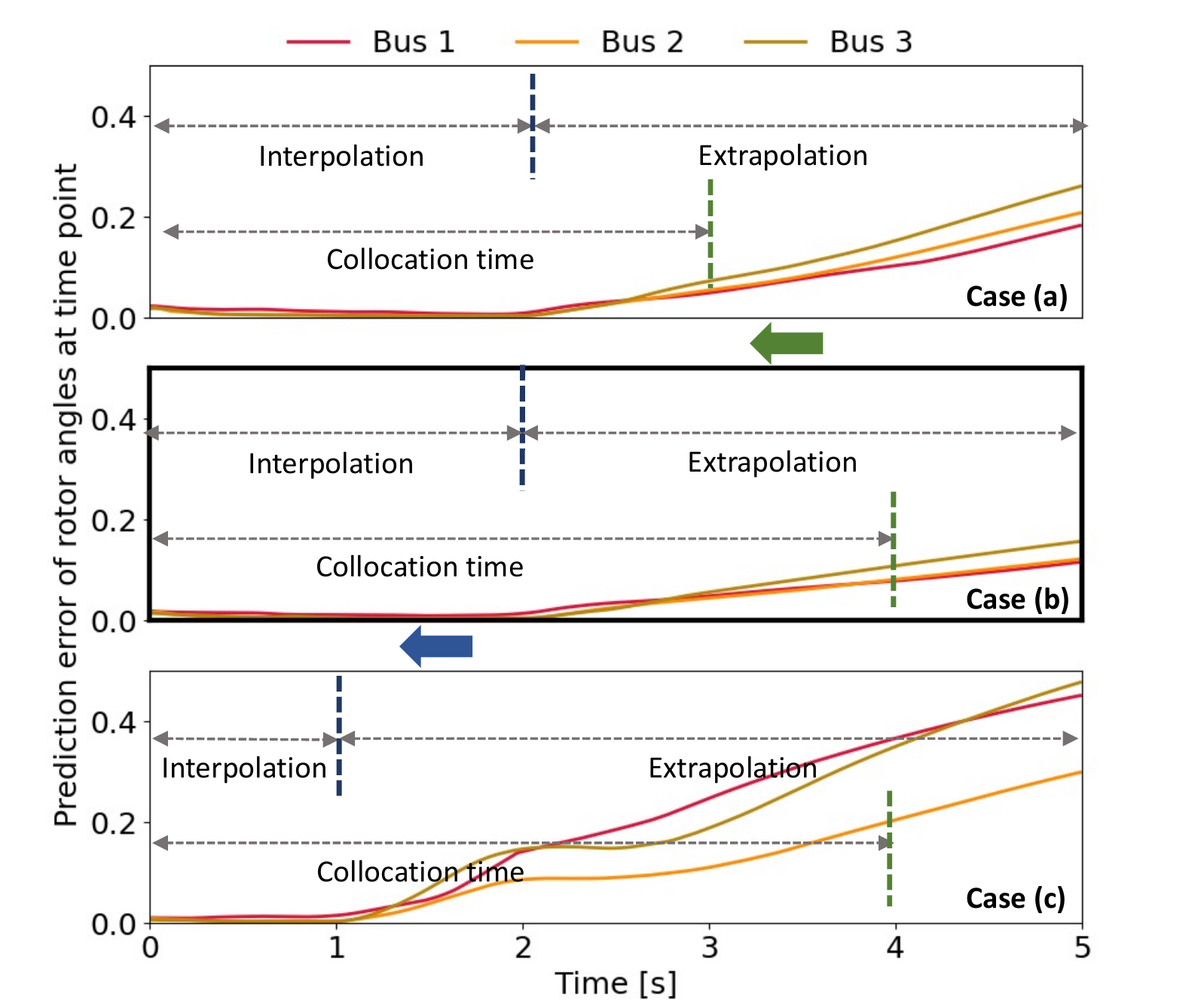}
    \caption{Prediction performance when varying the interpolation $T_{in}$ and extrapolation times $T_{end}$ for cases (a), (b) and (c).}
    \label{fig:fig4.4.3}
\end{figure}
This study investigated the SPINNs training (and sampling) strategies settings on the predictive performance. The user decided on training parameters, such as the length of the time-domain simulations (interpolation time $T_{in}$) and how to generate collocation points (without time-domain simulations). The generation of collocation points can be in interpolation times (so from $[0-T_{in}]$), and (or) extrapolation times (so from  $[T_{in}-T_{end}]$). $D1$, $D2$ are the number of generated points in interpolation, and extrapolation. Two studies were undertaken to unpack the combined impact of $T_{in},T_{end}$ and $D1$, $D2$ on the final predictive performance of SPINNs. 

\begin{figure}
\centering
     \begin{subfigure}[b]{.24\textwidth}
         \centering
         \includegraphics[width=.8\textwidth]{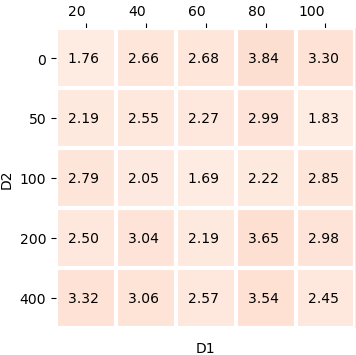}
        \caption{Interpolation error (\%)}
        \label{fig:heat_inter}
     \end{subfigure}
     \hfill
     \begin{subfigure}[b]{.24\textwidth}
        \centering
        \includegraphics[width=1\textwidth]{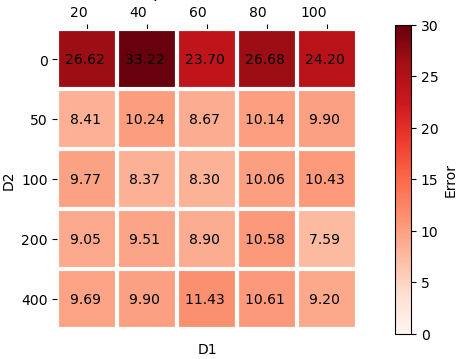}
        \caption{Extrapolation error (\%)}
        \label{fig:heat_extra}
    \end{subfigure}
\caption{The prediction errors (inter-, extrapolation) from SPINNs when varying the number of collocation points in the interpolation $D1$ and extrapolation $D2$ times.}
\label{fig:fig4.4.1}
\end{figure}

The first study investigated the impact of selecting the interpolation time $T_{in}$ and collocation time $T_{end}$. Three SPINN models were trained for the three cases (a) $T_{in}=2s$, $T_{end}=3s$, (b) $T_{in}=2s$, $T_{end}=4s$ (baseline), and (c) $T_{in}=1s$, $T_{end}=4s$. The prediction error was compared along the time span of $\SI{5}{\second}$ as shown in Fig.~\ref{fig:fig4.4.3}. 
The second study investigated balancing the selection of collection points in interpolation and extrapolation times $(D1,D2)$. The default interpolation time $[0-2]$s and prediction time $[0-4]$s were used. The results are in Fig.~\ref{fig:fig4.4.1}. 

The following analysis is for the two studies:
\begin{enumerate}[label=(\roman*)]
\item the first NN in the SPINN model approximated well the function between feature inputs and the state variables outputs, as the interpolation error was lower than the extrapolation error. Also, following the interpolation time, the prediction error started increasing as shown in Fig.~\ref{fig:fig4.4.3}. 
\item the extrapolation error was not evidently impacted by the number of $D1$ when the sampling interpolation time was fixed. However, the extrapolation error decreased by $17.8\%$ when the interpolation time was extended as in Fig.~\ref{fig:fig4.4.3} (b)-(c). Thus, the interpolation time length played a critical role in model's extrapolation performance. 
\item the extrapolation error largely reduced when the collocation points were considered in the training, as shown in Fig.~\ref{fig:fig4.4.1}. Once the regularisation was imposed, the extrapolation performance enhanced evidently. 
\item the collocation sampling time influenced the model performance. As shown in Fig.~\ref{fig:fig4.4.3}, the prediction error in time period $[3,5]$s was reduced by $8\%$ in case (a) compared with case (b) as additional collocation points were sampled in the extra time $[3,4]$s. 
\item SPINNs required $80\%$ less data than the baseline. This increase in efficiency results from the physical regularisation on the collocation points and the SPINN structure.
\end{enumerate}

This analysis demonstrated the key finding that SPINNs increase the prediction performance for longer dynamics without additional training data. The proposed SPINN model improved prediction performance and reduced the data dependency.

\subsection{Computational time}
The computational performance of SPINN was investigated, compared with a RK-45 ODE solver in online computation and also compared with PINNs in offline training time efficiency. 

\begin{table}
\centering
\caption{Real-time analysis of computations with SPINN and RK-45 ODE solver for different simulation time spans.}
\label{tab:tab2}
\begin{tabular}{|c|c|c|c|c|c|}
\hline
\multirow{2}{*}{} & \multicolumn{5}{c|}{Simulation times} \\ \cline{2-6} 
                  & 1s     & 2s     & 3s     & 4s     & 5s    \\ \hline
SPINN             & 0.04s  & 0.05s  & 0.05s  & 0.05s  & 0.06s \\ \hline
RK-45 solver      & 0.46s  & 0.50s  & 0.60s  & 0.61s  & 1.03s \\ \hline
\end{tabular}
\end{table}

The first study focused on the online computational times when the trained SPINNs would be used for prediction. The proposed SPINNs were compared against a representative numerical integration solver, the RK-45 ODE solver. The average results for computing/predicting a single OC are in Tab.~\ref{tab:tab2}. As the RK-45 solver required numerical integration in real-time the computational times were too high for real-time DSA for large systems, and when longer dynamics would like to be simulated. SPINNs had very low computational times of only $0.05$s making them promising for real-time dynamics study. This effect increased when simulating for longer times, e.g., in this example, SPINNs were $10$ times faster than RK-45 at $\SI{1}{\second}$ simulation time, and almost $20$ times faster at $\SI{5}{\second}$ simulations.

\begin{table}[t]
\centering
\scriptsize
\caption{The offline training time of the SPINN and PINNs when varying the number of seen OCs for training.}
\label{tab:tab4}
\begin{tabular}{|c|c|c|c|c|c|}
\hline
\multirow{2}{*}{}       & \multicolumn{5}{c|}{The number of seen OCs considered in training} \\ \cline{2-6} 
                        & 20          & 40          & 60          & 80          & 100        \\ \hline
SPINN & 11min 50s        & 12min 26s        & 12min 52s        & 13min 35s        & 14min 49s       \\ \hline
PINNs & 19min 51s       & 39min 48s       & 59min 30s       & 79min 13s       & 97min 16s      \\ \hline
\end{tabular}
\end{table}

The second study investigated offline training times. SPINN and PINN models were compared when using $100$ OCs. In PINNS, one PINN was trained for each of the $100$ OCs as the issue of PINNs is not considering simultaneously multiple OCs (Sec. \ref{sec3.1}) Conversely, the SPINN can consider multiple OCs in one training and be generalized to new OCs. As the results in Tab.~\ref{tab:tab4} demonstrate, SPINN reduced the offline training time by at most $90\%$, even without considering the time of computing and manually updating the physical model for PINNs training.

\section{Conclusions and future work}
In this paper, we have investigated the performance of the proposed SPINNs to predict power system dynamics under varying OCs. The inclusion of (a) a selected set of equations which are scalable to different OCs and (b) collocation points under unseen OCs and extrapolation time period, results in an increased prediction accuracy and a reduced labelled training data dependency, when compared with PINNs. With regard to the computation time, the SPINNs are largely faster than traditional simulators and needs less training time than PINNs.

There are still open questions and limits existing for future work. As the database was constructed synthetically with simulation tools, the patterns in the database may be constrained and not fully represent the real patterns of power system data. Additional analysis should be conducted to test the performance of the approach in the cases with more modals, more variance and in larger systems. The applicability of the approach to cascading failure events should also be tested. 

\bibliographystyle{IEEEtran}
\bibliography{ref.bib}

\end{document}